\documentclass[11pt]{article}

\usepackage[a4paper,margin=1in]{geometry}
\usepackage[utf8]{inputenc}
\usepackage[T1]{fontenc}
\usepackage{lmodern}
\usepackage{amsmath,amssymb,amsfonts}
\usepackage{graphicx}
\usepackage{textcomp}
\usepackage{booktabs}
\usepackage{tikz}
\usepackage{siunitx}
\usepackage{makecell}
\usepackage{microtype}

\usepackage[backend=biber, style=alphabetic,
            maxbibnames=9, minbibnames=1, giveninits=true,
            doi=false, isbn=false, url=false, eprint=false]{biblatex}
\DeclareSourcemap{\maps[datatype=bibtex]{\map{
  \step[fieldsource=entrykey, match=\regexp{\AANSI_AAMI_EC57_2012\Z}, final]
  \step[fieldset=label, fieldvalue={AAMI}]}}}
\AtEveryBibitem{\clearfield{volume}\clearfield{number}\clearfield{pages}\clearfield{month}}

\usepackage[colorlinks=true]{hyperref}

\newcommand{\keywords}[1]{\begin{quote}\small\textbf{Keywords:} #1\end{quote}}

\title{\textbf{Automated Fetal Magnetocardiography Reconstruction Using a Biophysical Forward Model}}
\date{}

\author{%
Jonas~Emrich$^{1}$,
Karin~Narushima$^{2}$,
Lena~Wunderl$^{3}$,
Philipp~Wunderl$^{3}$,\\
Tobias~Jensch$^{3}$,
Maximilian~Huber$^{3}$,
Peter~Ewert$^{2}$,
Peter~Fierlinger$^{3}$,\\
Annette~Wacker-Gussmann$^{2}$,
and~Reinhard~Heckel$^{1,4}$\thanks{This work was supported by Stiftung KinderHerz Deutschland gGmbH. Corresponding author: Reinhard Heckel (e-mail: reinhard.heckel@tum.de). Code is available at \url{https://github.com/MLI-lab/fetal_magnetocardiography}.}\\[8pt]
\normalsize $^{1}$School of Computation, Information and Technology, Technical University of Munich\\
\normalsize $^{2}$Clinic of Congenital Heart Disease and Pediatric Cardiology, German Heart Center Munich\\
\normalsize $^{3}$Physics Department E66, School of Natural Sciences, Technical University of Munich\\
\normalsize $^{4}$Munich Center for Machine Learning
}

\begin{document}
\maketitle

% Licence statement: the article is open access under CC BY 4.0 (confirmed via Crossref).
\begingroup\renewcommand{\thefootnote}{}\footnotetext{\copyright~2026 The Authors. This work is licensed under a Creative Commons Attribution 4.0 License. For more information, see \url{https://creativecommons.org/licenses/by/4.0/}. This paper appears in IEEE Transactions on Biomedical Engineering, DOI: \href{https://doi.org/10.1109/TBME.2026.3745242}{10.1109/TBME.2026.3745242}.}\endgroup

\begin{abstract}
    \textit{Objective:} Fetal magnetocardiography (fMCG) enables early detection of cardiac disease by non-invasively measuring fetal electrophysiological activity. 
    However, current signal estimation methods for fMCG require complex, labor-intensive human processing because they rely on blind source separation or template-based projection techniques.
    \textit{Methods:} We propose an automated signal reconstruction approach that recovers fetal and maternal cardiac activity from fMCG measurements without manual annotation. Our approach leverages the relationship between measurements and cardiac signals via a biophysically grounded forward model that represents cardiac sources as time-varying magnetic dipoles. This enables reconstruction of physiologically interpretable three-dimensional vectorcardiographic signals along with their spatial position.  
    \textit{Results:} Validation on synthetic data with known ground truth confirms accurate recovery of cardiac source parameters. On real fMCG recordings, the reconstruction shows strong agreement with established reference methods.
    \textit{Conclusion:} The proposed forward-model approach automatically reconstructs fetal and maternal cardiac activity from fMCG recordings and produces signals in an interpretable vectorcardiographic source domain.
    \textit{Significance:} Automated, interpretable fMCG reconstruction could make fetal cardiac assessment more reliable and support potential translation into prenatal care.
\end{abstract}


\section{Introduction}

In recent years, fetal Magnetocardiography (fMCG) has emerged as a viable tool for characterizing the cardiac activity of the fetal heart. It enables  the understanding of fetal arrhythmias, conduction abnormalities, and unexpected stillbirths~\cite{wacker-gussmannOPMbasedFetalMagnetocardiography2026,strandComplexNovelArrhythmias2020, wacker-gussmannContributionFetalMagnetocardiography2022, wacker-gussmannFetalMagnetocardiographyAlters2022, strandRepolarizationPredictorsFetal2020, strand2019LowCostFMCG, strand2019WaveformCharacteristics, strasburgerMagnetocardiographyFetalArrhythmias2008}. 

By measuring the magnetic field generated by the heart's electrical activity, fMCG enables a non-invasive assessment of the fetal cardiac electrophysiology. fMCG can provide diagnostic insights beyond conventional methods, particularly for high-risk cases, where traditional methods are unable to detect fetal arrhythmias that may otherwise remain unnoticed~\cite{wacker-gussmannFetalMagnetocardiographyAlters2022, wacker-gussmannContributionFetalMagnetocardiography2022}.

Detecting these magnetic fields requires highly sensitive magnetic field sensors because the signals are extremely weak, typically in the femtotesla to picotesla range. Recent advancements in sensor technology, particularly optically pumped magnetometers~\cite{hrenFetalMagnetocardiographyUsing2026}, have improved the feasibility of fMCG and Magnetocardiography (MCG) in general, and enabled cost-efficient, relatively small, person-sized magnetic shielded systems~\cite{wurmSmallScaleOptically2023, escalona-vargasCustomizedBedBased2025}. However, their reduced shielding and increased sensitivity to environmental noise demand more robust and accurate signal reconstruction.

In fMCG, a sensor array measures changes in the magnetic field from the fetal and maternal hearts, along with environmental noise. 
The aim is to reconstruct the underlying fetal cardiac signals from these noisy measurements, which are dominated by the stronger maternal heart signals.

\begin{figure*}[!t]
    \centering
    \includegraphics[width=.95\textwidth]{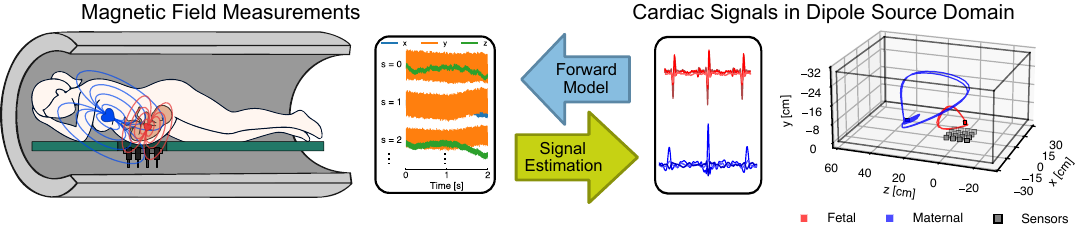}
    \caption{\textbf{Measurement setup and  cardiac source representation.} 
    The magnetic field is recorded by a sensor array under the abdomen while the patient lies in a magnetically shielded device. 
    The fetal and maternal cardiac signals are modeled as magnetic dipole sources, and the relationship to the measurements is described with a forward model. 
    The forward model is used to reconstruct fetal and maternal heart signals.}
    \label{fig:system}
\end{figure*}

Prior work on fetal signal extraction used blind source separation approaches, such as Independent Component Analysis (ICA), to decompose the measured field into source components~\cite{escalona-vargasCustomizedBedBased2025, wurmSmallScaleOptically2023, wacker-gussmannContributionFetalMagnetocardiography2022, strandComplexNovelArrhythmias2020, strandRepolarizationPredictorsFetal2020, murtaSegmentedIndependentComponent2015, yuMaternalMCGInterference2011, araujoFetalSourceExtraction2005}. 
These approaches assume statistical independence between fetal and maternal signals. This drives the decomposition into statistically rather than physically determined components and typically fragments each cardiac source across multiple components (see \hyperref[a:ica_limits]{Supplement~2} for illustration). 
As a result, the morphology of the extracted signals lacks a consistent physical interpretation and varies between recordings. This necessitates manual processing during reconstruction, annotating the extracted components as maternal, fetal, or noise, which is time-consuming and prone to error. 
Applying ICA across all spatial axes $(x,y,z)$ also ignores the explicit vectorial field information that could guide a biophysically grounded separation.
Other work applied spatial filtering methods~\cite{escalona-vargasComparisonFiveAlgorithms2018, wilsonSeparationPhysiologicalSignals2017,chen2006lmmse} to suppress or enhance signal components based on their alignment with spatial target structures. 
To construct signal and interference subspaces from noisy data, these approaches typically rely on manual template selection or on averaging spatial structures from hundreds to thousands of heartbeats, which neglects variation within and between beats.
However, neither approach explicitly models the relationship between the measured magnetic field and the underlying cardiac sources.

In this work, we propose a new reconstruction strategy for fMCG that uses a forward model relating the target cardiac signal to measured magnetic fields (see Fig.~\ref{fig:system}). This explicitly incorporates the spatial and vectorial structure of the observed magnetic field and enables fully automatic and accurate fMCG reconstruction. 

Our forward model represents each fetal and maternal cardiac source as a time-varying magnetic dipole, characterized by a time-dependent spatiotemporal dipole moment $\mathbf{m}(t)$ and spatial location $\mathbf{r}$. This captures the intrinsic cardiac dynamics in three spatial dimensions, analogous to vectorcardiography~\cite{FRANK}.

We estimate the source parameters by minimizing a regularized least-squares objective with a cross-correlation regularizer that supports separation between the recovered fetal and maternal sources.
This approach enables fully automated signal reconstruction and yields signals in an interpretable source domain that explicitly characterizes spatial position and spatiotemporal cardiac activity in three dimensions. This circumvents the limitations of prior signal-separation-based reconstruction methods. 

We first evaluate our approach on synthetic fMCG data and demonstrate that it accurately reconstructs the ground-truth dipole source parameters. In experiments using real fMCG data, the reconstruction approach yields estimates consistent with manual spatial-filtering-based processing. All results align with normative physiological values reported in the literature~\cite{wacker-gussmannOPMbasedFetalMagnetocardiography2026, strand2019WaveformCharacteristics}. 
These findings highlight the potential of our reconstruction approach for improving the practical feasibility of fMCG in clinical applications.

\subsubsection*{Related work}
Forward models that connect measurements to target signals, and thereby incorporate domain knowledge, are routinely used in other biomedical applications \cite{heckelDeepLearningComputational2025, westnerUnifiedViewBeamformers2022, bhatConstructionDynamicLead2021a}.
In MCG, the task is the imaging of adult cardiac activity at the heart level, typically requiring precise volumetric simulations and dense sensor arrays. Whereas the central challenge in fMCG lies in reconstructing ECG-like traces of the very weak fetal signal from mixtures dominated by maternal cardiac activity and noise.

In fMCG, dipole models have been explored to track fetal position and movement from already reconstructed fetal cardiac signals \cite{avciTrackingFetalMovement2018, EscalonaVargas2016GeneticDipoleLocalization}, but have not been leveraged for signal reconstruction.

\section{Methodology}

\begin{figure*}[tb]
    \centering
    \resizebox{.85\linewidth}{!}{\input{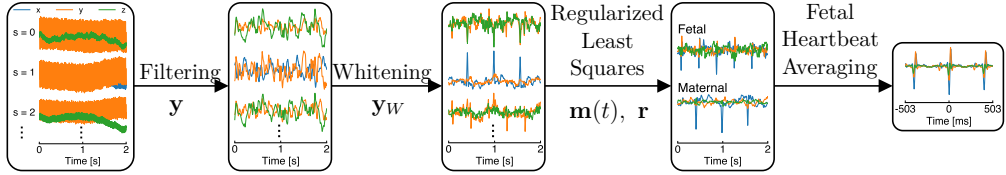}}
    \caption{\textbf{Forward model-based fMCG reconstruction.} Initial preprocessing steps enhance signal quality, after which cardiac activity in the equivalent dipole domain is estimated by solving a regularized least squares problem. Subsequent postprocessing prepares the results for clinical assessment. }
    \label{fig:pipeline}
\end{figure*}

In this section, we describe our forward model-based fMCG reconstruction pipeline, illustrated in Fig.~\ref{fig:pipeline}. 
We first preprocess the raw sensor measurement and then reconstruct by minimizing a regularized least-squares objective that relies on a forward model derived below. Finally, we perform postprocessing to prepare the reconstructions for clinical assessment. 

\subsection{Data acquisition}\label{ssec:setup}
We use the fMCG system with optically pumped magnetometers at the Department of Congenital Heart Disease, TUM University Hospital, German Heart Center Munich~\cite{wurmSmallScaleOptically2023}. Recordings were obtained with the informed consent of participants and with approval from the Ethics Committee of the Technical University of Munich, in accordance with the Declaration of Helsinki.

The device consists of a cylindrical, person-sized magnetic shield enclosing a movable bed positioned above a planar $4 \times 4$ sensor array of sixteen zero-field magnetometers with a \SI{4}{cm} spacing (see Fig.~\ref{fig:system} left panel). The sensors measure two of the three magnetic field directions ($x, y, z$), with adjacent sensors rotated by \ang{90} to cover all components. 

Data is recorded with a sampling frequency of \SI{1}{kHz} and downsampled to \SI{250}{Hz} after preprocessing to reduce computational complexity in solving the inverse problem.

The resulting measurements for each channel $c \in \{x,y,z\}$ and sensor $s \in \{0,1,\dots,S-1\}$ are $y_{s,c}(t)$ for time $t \in \{0,1,\dots,T-1\}$.

\subsection{Preprocessing}
We first preprocess the data with the following four steps, applied sequentially.

\subsubsection{Frequency filtering}
The raw recordings are first bandpass filtered between \SI{1}{Hz} to \SI{40}{Hz} using a $4$th-order zero-phase forward-backward Butterworth filter to suppress baseline wander and environmental noise. In addition, notch filters at frequencies $f \in \{16.7, 38, 50, 76\}$\si{Hz} are applied to attenuate narrowband interferences from known environmental sources, including power line noise, tram power line noise, and internal noise from the optically pumped magnetometers.

\subsubsection{Removing faulty sensor channels}
Faulty sensor channels exhibiting an unphysical signal range or outlying variability are excluded before processing (details in \hyperref[supp:preprocessing]{Supplement~1}).

\subsubsection{Removing temporal artifacts}
Segments of temporal outliers or artifacts are removed using  multi-resolution thresholding in the wavelet domain~\cite{strasserMotionArtifactRemoval2012} (see \hyperref[supp:preprocessing]{Supplement~1}). The recordings are then divided into artifact-free segments of at least 10 seconds, excluding periods with substantial motion artifacts or magnetic interference.

\subsubsection{Pre-whitening}
In multivariate magnetic field recordings, the spatial structure of signals and sources captured by the planar sensor array can be leveraged to suppress noise via pre-whitening. Given the raw field data $\mathbf{y}(t) \in \mathbb{R}^{2S}$ recorded by $S$ dual-axis sensors, the whitened data is obtained as 
\begin{equation}
    \mathbf{y}_W(t) = \mathbf{W} \, \mathbf{y}(t),
\end{equation}
where $\mathbf{W} = \mathbf{S}^{-1/2}\mathbf{U}^\top
\in \mathbb{R}^{2S \times 2S}
$ is the whitening matrix derived from the noise covariance $\mathbf{\Sigma}_N \in \mathbb{R}^{2S \times 2S}$ estimated from an empty-device measurement. Here, $\mathbf{\Sigma}_N = \mathbf{U}\mathbf{S}\mathbf{U}^\top$ is the singular value decomposition. 
This approach attenuates signal components aligned with the noise subspace while preserving those orthogonal to it, thereby suppressing environmental interference and physiological artifacts.

\subsection{Forward model}
We now formulate a forward model that relates the measured magnetic fields to the fetal and maternal cardiac signals. 

Each cardiac source is modeled as a magnetic dipole, defined by its position $\mathbf{r}$ and magnetic dipole moment $\mathbf{m}(t)$, which describes the spatiotemporal strength and direction of cardiac activity. In contrast to electric current dipole models, magnetic dipoles do not depend on assumptions about the volume conductor's geometry and properties. The magnetic field measured outside the maternal abdomen is well represented by an equivalent magnetic dipole~\cite{Stinstra2001, avciTrackingFetalMovement2018}. This makes magnetic dipoles a practical choice for modeling the contributions from fetal and maternal sources.

The magnetic flux density $\mathbf{B} \in \mathbb{R}^{3}$ at sensor position $\mathbf{r}_s \in \mathbb{R}^{3}$ produced by a magnetic dipole $d$ with moment $\mathbf{m}_d \in \mathbb{R}^3$ located at $\mathbf{r}_d \in \mathbb{R}^3$ is given by
\begin{equation}\label{eq:single_dipole}
    \mathbf{B}(\mathbf{m}_d, \mathbf{r}_\Delta) = \underbrace{\frac{\mu}{4\pi}\left[\frac{3\mathbf{r}_\Delta \mathbf{r}_\Delta^\top}{\Vert\mathbf{r}_\Delta\Vert^{5}} - \frac{\mathbf{I}}{\Vert\mathbf{r}_\Delta\Vert^{3}}\right]}_{\mathbf{A}(\mathbf{r}_d, \mathbf{r}_s)}\mathbf{m}_d ,
\end{equation}
where $\mathbf{r}_\Delta = \mathbf{r}_s - \mathbf{r}_d$ is the vector from dipole $d$ to the observation point $s$, $\mu$ is the magnetic permeability and $\mathbf{I}$ is the identity matrix. For a fixed sensor position $\mathbf{r}_s$, the lead field matrix $\mathbf{A}(\mathbf{r}_d, \mathbf{r}_s)$ depends only on the dipole position.
Equation~\eqref{eq:single_dipole} is non-linear in $\mathbf{r}$, but linear in $\mathbf{m}$, acting as a linear mapping from the cardiac signal in dipole space to the measured field domain.

Because biological tissues are largely non-magnetic \cite{schenckPhysicalInteractionsStatic2005}, magnetic field propagation through the human body closely resembles that in free space, and the permeability is therefore approximately equal to that of free space $\mu \approx \mu_0$.

We define the time-varying fetal and maternal magnetic dipole moment vectors as $\mathbf{m}_f(t), \mathbf{m}_m(t) \in \mathbb{R}^{3}$ and their positions $\mathbf{r}_f, \mathbf{r}_m \in \mathbb{R}^{3}$ along with the stacked parameter vectors:
\begin{equation}
    \mathbf{m}(t) = \begin{bmatrix} \mathbf{m}_f(t) \\ \mathbf{m}_m(t) \end{bmatrix} \in \mathbb{R}^{6}
    \qquad
    \mathbf{r} = \begin{bmatrix} \mathbf{r}_f \\ \mathbf{r}_m \end{bmatrix} \in \mathbb{R}^{6}.
\end{equation}
The forward model for fMCG is then
\begin{equation}\label{eq:forward_model}\setlength{\arraycolsep}{1pt}
    \underbrace{\begin{bmatrix} \mathbf{y}_{1}(t) \\ \vdots \\ \mathbf{y}_{S}(t) \end{bmatrix}}_{\mathbf{y}(t) \in \mathbb{R}^{2S}}
    = \mathbf{M}
    \underbrace{\begin{bmatrix}
        \mathbf{A}(\mathbf{r}_{f}, \mathbf{r}_{1}) & \mathbf{A}(\mathbf{r}_{m}, \mathbf{r}_{1}) \\
        \vdots & \vdots \\
        \mathbf{A}(\mathbf{r}_{f}, \mathbf{r}_{S}) & \mathbf{A}(\mathbf{r}_{m}, \mathbf{r}_{S})
    \end{bmatrix}}_{\mathbf{F}(\mathbf{r}) \in \mathbb{R}^{3S\times6}}
    \underbrace{\begin{bmatrix} \mathbf{m}_{f}(t) \\ \mathbf{m}_{m}(t) \end{bmatrix}}_{\mathbf{m}(t) \in \mathbb{R}^{6}}
    + \underbrace{\begin{bmatrix} \mathbf{n}_{1}(t) \\ \vdots \\ \mathbf{n}_{S}(t) \end{bmatrix}}_{\mathbf{n}(t) \in \mathbb{R}^{2S}},
\end{equation}
where $\mathbf{y} \in \mathbb{R}^{2S}$ is the observed field at $S$ sensors with two spatial directions each, $\mathbf{n} \in \mathbb{R}^{2S}$ is interfering noise, and $\mathbf{M} \in \mathbb{R}^{2S \times 3S}$ masks the unobserved axis per sensor.
The forward operator $\mathbf{F} \in \mathbb{R}^{3S\times6}$ forms a block matrix, in which each $3 \times 3$ block $\mathbf{A}(\mathbf{r}_d, \mathbf{r}_s)$ maps from a dipole at $\mathbf{r}_d$ to a sensor at $\mathbf{r}_s$, as in Equation~\eqref{eq:single_dipole}.
Since the sensor locations $\mathbf{r}_1, \dots, \mathbf{r}_S$ are known, we write the forward operator as $\mathbf{F}(\mathbf{r})$, omitting the sensor locations for brevity.
This model gives a unified, low-dimensional representation in which the dipole moments $\mathbf{m}_f(t)$ and $\mathbf{m}_m(t)$ form cardiac vector loops analogous to vectorcardiography (see Section~\ref{ssec:vectorial}), while $\mathbf{r}_f$ and $\mathbf{r}_m$ describe their spatial locations.
Reconstruction then corresponds to estimating the cardiac source parameters $\mathbf{m}(t)$ and $\mathbf{r}$ from noisy observations $\mathbf{y}$. 
 
Although our model approximates each heart as an equivalent dipole with fixed position, the moment $\mathbf{m}(t)$ remains a full three-dimensional, time-varying vector that captures changes in spatial orientation. This contrasts with ICA or template-based spatial filters, which assume a single spatial pattern per source, fixed across all beats. 
The fixed-position assumption is most appropriate for short time windows, where approximate stationarity holds. ICA and spatial filtering incorporate the same assumption, relying on stationary signal statistics. 

Our formulation extends naturally to moving dipoles or more sources, at the cost of a harder inverse problem, higher computational cost, and stronger regularization constraints to ensure stable reconstructions. We explore an extension to time-varying positions in \hyperref[a:time_varying_positions]{Supplement~5}.

Higher-order poles, such as quadrupoles or octupoles, decay rapidly with distance, while the dipole dominates the multipole expansion, especially when the source is small relative to the observation distance. Dipole models are therefore widely used in biomedical source modeling \cite{westnerUnifiedViewBeamformers2022, bhatConstructionDynamicLead2021a, avciTrackingFetalMovement2018}, offering a simple and effective representation.

\subsection{Signal reconstruction with regularized least-squares}
We estimate the dipole parameters, i.e., the magnetic moment vector $\mathbf{m}(t)$ and the position $\mathbf{r}$, from the field measurements $\mathbf{y}(t)$ by minimizing
\begin{equation}\label{eq:loss}
    \mathcal{L}(\mathbf{m}(t), \mathbf{r}) = \mathcal{D}(\mathbf{m}(t), \mathbf{r}) + \mathcal{R}(\mathbf{m}(t)),
\end{equation}
which combines a data-consistency term
\begin{equation}\label{eq:datacons}
    \mathcal{D}(\mathbf{m}(t), \mathbf{r}) = \frac{1}{2}\,\frac{\sum_{t}\left\Vert\mathbf{y}_W(t) - \mathbf{W}\mathbf{M}\mathbf{F}(\mathbf{r})\mathbf{m}(t)\right\Vert^2_2}{\sum_{t}\left\Vert\mathbf{y}_W(t)\right\Vert^2_2}
\end{equation}
with a regularizer $\mathcal{R}$ on the moment trajectory. The data-consistency term enforces agreement between the whitened magnetic field observation $\mathbf{y}_W(t)$ and the whitened forward prediction $\mathbf{W}\mathbf{M}\mathbf{F}(\mathbf{r})\mathbf{m}(t)$. Normalization ensures comparable scales across configurations with varying signal power.

\subsubsection{Regularization}
Although the solution to the inverse problem can be estimated by minimizing the data-consistency term alone, we observed crosstalk between fetal and maternal cardiac activity in some reconstructed dipole moments, potentially due to variations in signal quality, model mismatches, or whitening.

To improve reconstruction robustness, we introduce a cross-correlation regularizer $\mathcal{R}$ to promote separability between the reconstructed fetal and maternal source signals. 
Following second-order blind identification (SOBI)~\cite{SOBI}, this term minimizes the cross-correlation between the estimated fetal and maternal dipole moments across multiple time lags $\tau \in [-\ell, \dots, \ell]$.
The penalty is computed over $K$ overlapping windows $k = 0, \dots, K-1$ (of length $N_k$ with \SI{50}{\percent} overlap) as 
\begin{equation}
    \mathcal{R}(\mathbf{m}(t)) = \lambda \, \sqrt{ \frac{1}{9K(2\ell+1)} \sum_{\tau=-\ell}^\ell \sum_{k=0}^{K-1} \left\Vert\boldsymbol{C}_k(\tau)\right\Vert^2_F },
\end{equation}
where $\lambda$ denotes the regularization strength, $\Vert\cdot\Vert^2_F$ is the Frobenius norm, and the factor 9 accounts for the number of matrix entries.
The normalized cross-correlation matrix $\boldsymbol{C}_k(\tau) \in \mathbb{R}^{3 \times 3}$ between fetal and maternal dipole moments is 
\begin{equation}
    \left[\boldsymbol{C}_k(\tau)\right]_{ij} = \frac{\left[\mathbf{R}_k(\tau)\right]_{ij}}{\sigma^{(k)}_{f,i} \,\sigma^{(k)}_{m,j}}, \quad i,j \in \{x,y,z\},
\end{equation}
with the zero-padded sample cross-covariance
\begin{equation}
    \mathbf{R}_k(\tau) = \frac{1}{N_k}\sum_n \tilde{\mathbf{m}}_{f}^{(k)}(n) \, \tilde{\mathbf{m}}_{m}^{(k)\top}(n-\tau).
\end{equation}
Here $\tilde{\mathbf{m}}_{f}^{(k)} \in \mathbb{R}^3$ and $\tilde{\mathbf{m}}_{m}^{(k)}$ are the mean-centered fetal and maternal dipole moments in window $k$ (zero-padded outside $[1, N_k]$), and $\sigma^{(k)}_{f,i}$ and $\sigma^{(k)}_{m,j}$ are their component-wise standard deviations.
A comparative example illustrating the effect of regularization on the dipole moment reconstruction can be found in \hyperref[a:regularization]{Supplement~3}.

\subsubsection{Optimization}
We minimize the objective in Equation~\eqref{eq:loss} via gradient-based optimization using the Lamb optimizer~\cite{You2020LAMB}, a variant of the widely established Adam optimizer~\cite{Kingma2015Adam} that incorporates adaptive learning rates through parameter groupwise normalization. This compensates for differences in magnitude across fetal and maternal moment and position parameters.

Dipole positions are initialized within physiologically plausible regions and constrained via bounded reparameterization to remain within the device. The magnetic moments are initialized via a strongly regularized pseudo-inverse estimate using the whitened forward model. Additional details are given in \hyperref[a:optimization]{Supplement~6}.

\subsection{Selective Averaging}\label{ssec:selective_averaging}

After optimization, the reconstructed magnetic dipole moments offer a continuous, three-dimensional characterization of fetal and maternal cardiac activity throughout the recording.
For detailed morphological assessment, such as waveform annotation, fMCG processing usually applies selective heartbeat averaging over beats with similar heart rate to enhance morphological details.
This trades beat-to-beat variability for signal quality, but is an optional post-processing step that enables cardiac interval annotation under noisy conditions.

The reconstructed signal is segmented into heartbeats by aligning R-peak positions. We use the visibility-graph-based R-peak detector~\cite{emrich2023vg} as it provides fast and sample-accurate beat detection. 
Then, consecutive segments of heartbeats are averaged when the fetal heart rate is near baseline, defined as within 5 bpm of a stable minimum during fetal quiescence~\cite{strand2019WaveformCharacteristics}. Fetal movement is evaluated using fMCG actography~\cite{lutter2011actography}. Selection criteria may be adjusted according to the analysis.

This produces a unified, multi-axis visualization of waveform morphology that supports the assessment of cardiac intervals (see Section~\ref{ssec:reconstruction}).

\section{Results}\label{sec:results}
We evaluate the forward model-based reconstruction approach against two established methods. Splined Independent Component Subtraction (splined ICS)~\cite{yuMaternalMCGInterference2011}, an ICA-based method, that reconstructs the maternal signal from annotated components and subtracts it from the raw data. The minimum norm covariance projector~\cite{wilsonSeparationPhysiologicalSignals2017, chen2006lmmse}, referred to as spatial filtering in the following, constructs a spatial template of the fetal heartbeat by averaging the raw data at fetal beat positions obtained from ICA-annotated reference. 
Both baselines require manual intervention, and are commonly used in fMCG analyses~\cite{wacker-gussmannOPMbasedFetalMagnetocardiography2026,strandComplexNovelArrhythmias2020, wacker-gussmannContributionFetalMagnetocardiography2022, wacker-gussmannFetalMagnetocardiographyAlters2022, strandRepolarizationPredictorsFetal2020, strand2019LowCostFMCG, strand2019WaveformCharacteristics, strasburgerMagnetocardiographyFetalArrhythmias2008}.

Since ground truth measurements are unavailable in fMCG, reconstruction performance is first assessed on synthetic data with known ground truth (see Section~\ref{ssec:synthetic}). Clinical validity is then evaluated on real fMCG data by direct comparison with spatial filtering, an established approach in fMCG processing~\cite{strand2019LowCostFMCG, wacker-gussmannContributionFetalMagnetocardiography2022, escalona-vargasComparisonFiveAlgorithms2018, escalona-vargasCustomizedBedBased2025}.

We find that on synthetic data, the forward model reconstructs both sources with low error, and on real data, its fully automated reconstruction agrees with spatial filtering and yields cardiac intervals consistent with reported ranges.

All experiments use $500$ optimization iterations with parameter-specific learning rates. Complete settings are detailed in \hyperref[a:optimization]{Supplement~6}, where reasonable variations give similar results. Complete processing of a \SI{5}{min} recording takes under \SI{1}{min} with GPU acceleration on an NVIDIA Quadro RTX 6000.

\subsection{Validation on synthetic data}\label{ssec:synthetic}

To evaluate reconstruction performance with known ground truth, realistic synthetic fMCG data were generated using the dipole model. The forward model recovered both fetal and maternal sources directly in the dipole domain while achieving field-domain reconstruction errors on par with spatial filtering and below those of splined ICS. Sources are simulated as static, so this tests identifiability of source parameters rather than validating the constant position assumption. \hyperref[a:time_varying_positions]{Supplement~5} examines movable dipoles and indicates that the static dipole is a reliable trade-off when positions vary slowly.

Cardiac activity of fetal and maternal heart was simulated with dipole moments derived from real vectorcardiographic data~\cite{BousseljotKreiselerSchnabel1995CARDIODAT}. These vectorcardiographic signals are scaled to realistic dipole magnitudes with $\pm 20\%$ random variation, randomly rotated, and placed at realistic positions with \SI{2}{cm} standard deviation between simulated recordings. Noise from a real empty device recording was added at \SI{0}{dB} signal-to-noise ratio, combined with \SI{5}{\percent} Gaussian noise.

Reconstruction performance in the field domain is evaluated using the R-peak-normalized root-mean-square error (NRMSE). As the forward model also provides estimates of the source parameters, dipole moments are evaluated using NRMSE, and dipole positions using root-mean-square error (RMSE) with respect to the simulation ground truth. Because the spatial filter exhibits an SNR dependent typically sub-unity gain~\cite{wilsonSeparationPhysiologicalSignals2017}, the evaluation metrics are scale-adjusted to ensure a fair comparison of the waveform reconstruction.

A total of 30 fMCG measurements were simulated. ICA components were manually annotated to enable reconstruction with ICS and spatial filtering. 

\begin{figure}[!t]
    \centering
    \includegraphics[width=0.6\linewidth]{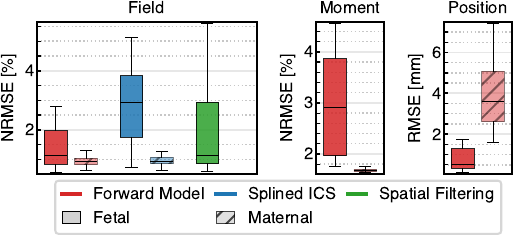}
    \caption{\textbf{Synthetic reconstruction performance.} 
     Reconstruction errors across $N=30$ simulated trials for fetal (solid) and maternal (hatched). The forward model (red) achieves field NRMSE on par with spatial filtering and below splined ICS, and is the only method that provides dipole moment and position estimates.}
    \label{fig:synthetic_fit}
\end{figure}

Reconstruction errors across all synthetic measurements are shown in Fig.~\ref{fig:synthetic_fit}. All methods accurately reconstruct the simulated fetal and maternal signals, yielding low errors in the field domain, with the forward model and spatial filtering achieving the smallest errors. The maternal signal is inherently reconstructed by the forward model, and splined ICS also yields a maternal estimate before it internally subtracts its contribution. 
In contrast, the covariance projection spatial filter does not directly yield a simultaneous maternal estimate, as it applies a sub-unity gain rather than a projection, and its residual remains fetal-contaminated~\cite{wilsonSeparationPhysiologicalSignals2017}.
Notably, the forward model additionally provides reconstructions of the dipole moments and positions, which also exhibit low errors. Maternal errors are smaller for the field and dipole domains, while the position error is slightly larger, reflecting the easier reconstruction in amplitude but slightly more difficult localization due to the field patterns originating from farther away.

\subsection{Reconstructing cardiac dipole activity}\label{ssec:reconstruction}

We now show representative real-world fMCG reconstructions to illustrate the estimated cardiac signals. A qualitative comparison of reconstructed heartbeat morphologies indicates that the forward model recovers physiologically plausible fetal cardiac waveforms comparable to those from splined ICS and spatial filtering. 

\begin{figure}[!t]
    \centering
    \includegraphics[width=0.6\linewidth]{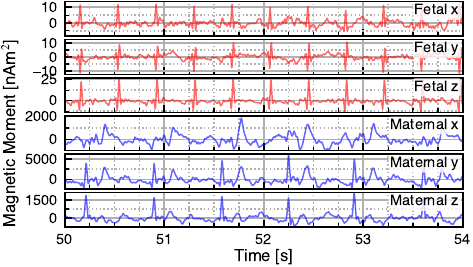}
    \caption{\textbf{Reconstructed magnetic dipole moments.} A four-second excerpt of reconstructed fetal and maternal magnetic dipole moments, illustrating the recovered cardiac characteristics over time.}
    \label{fig:m_time}
\end{figure}

Our forward model-based approach provides reconstructions in the dipole source domain, whereas conventional fMCG reconstruction methods yield signals only in the field domain.
Fig.~\ref{fig:m_time} shows an example of the reconstructed magnetic dipole moment over a four-second interval, showing cardiac activity from both the fetus and the mother. The fetal signal is two orders of magnitude weaker than the maternal signal, as expected due to its smaller size and lower electrophysiological amplitude. 
The method yields a dipole time series spanning the entire recording, allowing thorough physiological inspection analogous to Holter monitoring.

Signals in the measured field domain are obtained by projecting the estimated dipole parameters through the forward model. In contrast to ICS and spatial filtering, the forward map $\mathbf{F}$ enables prediction of the unobserved third field direction at each sensor, yielding the full set of $3S$ channels.

\begin{figure*}[!t]
    \centering
    \includegraphics[width=\linewidth]{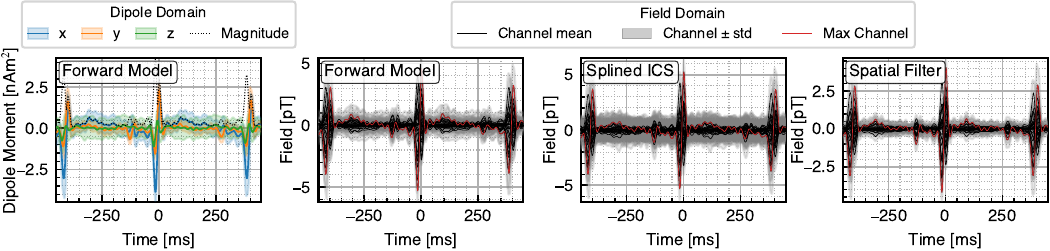}
    \caption{\textbf{Comparison of fMCG reconstructions in dipole and field domain.} The forward model-based reconstructions of the fetal source in the dipole and field representation are compared against reconstructions with spatial filtering and ICA-based techniques. Plots show the selectively averaged heartbeat waveforms and their one-standard-deviation margins (std). The forward model is the only method providing reconstructions in the source domain.}
    \label{fig:domains}
\end{figure*}

Fig.~\ref{fig:domains} illustrates the relationship between the dipole and the field domain reconstructions, alongside the corresponding field domain results of splined ICS and spatial filtering. Each plot displays a window of three consecutive heartbeat waveforms obtained via selective averaging. The dipole domain shows averages across all three spatial axes, along with their one-standard-deviation margins and the magnitude. In the field domain, butterfly plots are typically used to display all channel averages overlaid. In addition to the one-standard-deviation margins, the mean channel with the largest signal energy is highlighted for reference. 

All representations show the prominent waveform features of the fetal cardiac cycle, with similar waveform morphologies for the P-wave, QRS complex, and T-wave. In healthy fetuses, the T-wave is often flat and hardly visible~\cite{strand2019WaveformCharacteristics}. The averaged-beat standard deviation is lowest for the spatial filter, as it suppresses signal components that do not align with the fetal template. The forward model also shows low variability, whereas splined ICS exhibits the highest variability, as it primarily removes maternal signals, leaving noise or residual components.
The spatial filter output has a lower amplitude, reflecting its SNR-dependent gain and that it does not produce physical values, in contrast to the forward model or ICS.

\subsection{Fetal cardiac time intervals} \label{ssec:cardiac_intervals}

After demonstrating that the forward model recovers ground truth in synthetic data (see Section~\ref{ssec:synthetic}), we now evaluate it on real fMCG recordings, where no ground truth exists. We compare it against spatial filtering-based reconstruction on a typical downstream analysis task: measuring cardiac waveform intervals. In our experiments, both methods agree closely with intraclass correlation coefficients exceeding $0.9$, and mean absolute differences below typical inter-annotator variability reported for manual ECG analysis. While spatial filtering is a reference method, not ground truth, this close agreement shows consistency with an established method.

A dataset of $N=31$ fMCG measurements from 22 healthy subjects is used, with gestational ages ranging from $21$ to $40$ weeks (mean $\pm$ std: $31.1 \pm 4.98$), including 9 from the second trimester. For a fair comparison, we apply identical preprocessing to both methods and average the same heartbeats selected as described in Section~\ref{ssec:selective_averaging}. 

Cardiac waveforms were manually measured in the field domain, as spatial filtering produces results only in this domain. While annotation in the dipole domain is also possible and may offer physiological benefits by providing three condensed, interpretable leads in the source domain, this study focused on the field domain to allow direct comparison with spatial filtering. 

The standard cardiac intervals, including P, PR, QRS, QT, and RR intervals, were measured following~\cite{wacker-gussmannOPMbasedFetalMagnetocardiography2026, strand2019WaveformCharacteristics}, with QTc from Bazett's correction $\text{QTc} = \text{QT}/\sqrt{\text{RR}}$. We annotated both reconstructions side by side to control for annotation inconsistencies and isolate reconstruction differences.

All measured intervals for both methods fall within the \SI{95}{\percent} prediction intervals reported for healthy fetuses (mean values: RR \SI{429}{ms}, P \SI{50}{ms}, PR \SI{108}{ms}, QRS \SI{57}{ms}, QT \SI{254}{ms}, QTc \SI{388}{ms}, all within \cite{wacker-gussmannOPMbasedFetalMagnetocardiography2026, strand2019WaveformCharacteristics}), supporting the physiological plausibility of both reconstruction approaches.

The agreement between the two approaches is reported in Table~\ref{tab:reconstruction_diffs}. Mean absolute differences are low across all intervals, with higher values observed for subtle waves such as the P-wave and lower values for prominent features such as the R-peak. 
Mean absolute differences across all intervals, ranging from $1.85$ to $6.36$~\si{ms}, are below the inter-annotator variability of manual ECG analysis ($7 - 19$~\si{ms} IQR)~\cite{Boeck2023}. The difference due to the reconstruction method is therefore smaller than the variability of manual cardiac waveform interpretation. In \hyperref[a:annotation_example]{Supplement~4}, we provide interval definitions and investigate differences across gestational age, but observe no relevant trend.

\begin{table}[!t]
    \centering
    \caption{Cardiac Interval Comparison Between Forward Model and Spatial Filtering}
    \label{tab:reconstruction_diffs}
    \setlength{\tabcolsep}{4pt}
    \footnotesize
    \begin{tabular}{l|cccccc}
    \toprule
     & RR & P & PR & QRS & QT & QTc \\
    \midrule
    Rel. diff. (\%)       & 0.43 & 4.57 & 2.34 & 3.09 & 1.61 & 1.66 \\
    Abs. diff. (ms)       & \makecell{1.85\\$\pm$1.50} & \makecell{2.29\\$\pm$1.44} & \makecell{2.52\\$\pm$1.52} & \makecell{1.78\\$\pm$1.58} & \makecell{4.00\\$\pm$4.24} & \makecell{6.36\\$\pm$6.23} \\ \midrule
    Expert IQR (ms)       & -- & 16.0 & 19.0 & 7.0 & 15.0 & -- \\
    \bottomrule
    \end{tabular}
    \\[2pt]
    {\footnotesize\raggedright Relative differences are means, absolute differences mean $\pm$ std ($N=31$). Expert IQR: inter-annotator variability for manual ECG analysis~\cite{Boeck2023}.\par}
\end{table}

The agreement between spatial filtering-derived intervals and forward model reconstructions is quantified by intraclass correlation coefficients (ICC) reported in Table~\ref{tab:icc_ica_field}. A subset of $N=22$ recordings from independent patients (gestational age: $26-40$, mean $\pm$ std: $32.5 \pm 4.42$) was used to ensure statistical independence, removing 9 repeated recordings from patients measured at several gestational ages. All ICC values exceed $0.9$, indicating strong overall agreement, with subtle features such as P-wave intervals exhibiting slightly lower but still strong agreement. These results support that the forward model reliably reconstructs fetal cardiac signals comparable to those of spatial filtering-based processing.

\begin{table}[!t]
    \centering
    \caption{Agreement of spatial filtering and forward-model}
    \label{tab:icc_ica_field}
    \setlength{\tabcolsep}{6pt}
    \footnotesize
    \begin{tabular}{l|cccccc}
        \toprule
        Interval & RR & P & PR & QRS & QT & QTc \\
        \midrule
        ICC(3,1) & 0.996 & 0.908 & 0.973 & 0.914 & 0.977 & 0.973 \\
        95\% CI & \makecell{0.990\\1.000} & \makecell{0.790\\0.960} & \makecell{0.940\\0.990} & \makecell{0.810\\0.960} & \makecell{0.940\\0.990} & \makecell{0.930\\0.990} \\
        \bottomrule
    \end{tabular}
    \\[2pt]
    {\footnotesize\raggedright ICC(3,1): two-way mixed-effects, absolute-agreement intraclass correlation between spatial-filtering and forward-model reconstructions ($N=22$). CI shown as lower/upper bound.\par}
\end{table}

\subsection{Spatiotemporal cardiac representation}\label{ssec:vectorial}
Our method reconstructs the fetal cardiac source as a three-dimensional vector $\mathbf{m}_f(t)$, providing a representation of the intrinsic cardiac activity that other methods do not provide. This is analogous to vectorcardiography~\cite{FRANK}, where three orthogonal ECG leads capture the heart's electrical dynamics as a spatial trajectory. This unified vector representation provides spatiotemporal information essential for robust and standardized fetal cardiac analysis~\cite{fECG_standard_2020,horigomeStandardizationPQRSTWaveform2006}.

\begin{figure}[!t]
    \centering
    \includegraphics[width=0.6\linewidth]{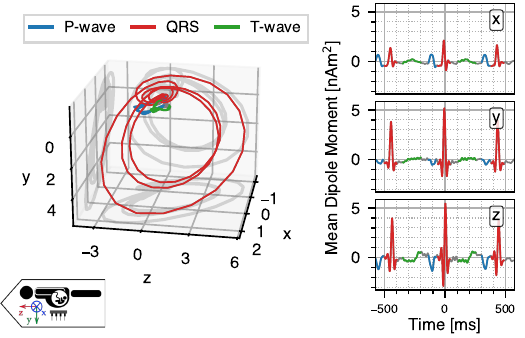}
    \caption{\textbf{Spatiotemporal cardiac source representation.} Reconstructed fetal cardiac vector loop with automatically segmented cardiac waves, illustrating the three-dimensional cardiac cycle. Corresponding dipole moment components along the x, y, and z axes.}
    \label{fig:delineation}
\end{figure}

Fig.~\ref{fig:delineation} shows the reconstructed fetal magnetic moment $\boldsymbol{m}_f(t)$ forming a vector loop over the cardiac cycle, analogous to vectorcardiography.
Waveform morphology strongly depends on the dipole axis. In this example (see Fig.~\ref{fig:delineation}), the fetal P-wave is prominent along the x- and z-axis but weak in the y-axis. The T-wave, which is known to be difficult to detect in fMCG \cite{strand2019WaveformCharacteristics}, is nearly absent and only slightly visible along the z-axis. This axis-dependence arises from the three-dimensional nature of cardiac dipole activity. Changes in the spatial orientation and rotation of the vector loop yield different projections onto each fMCG axis, producing different per-axis morphologies. This is inherent to fMCG, because fetal position and rotation naturally vary between measurements.

\subsection{Automated cardiac delineation based on vectorcardiographic representation}
To demonstrate the utility of the three-dimensional vectorcardiographic representation produced by our forward model, we assess its feasibility for automated cardiac delineation. Cardiac delineation segments the heartbeats into their individual waves (P, QRS, T), which is essential for quantitative analysis of cardiac intervals.

Due to the scarcity of clean and annotated fMCG data, training a delineation model directly on fMCG is infeasible. Instead, we demonstrate that the three-dimensional dipole representation enables transfer learning from vectorcardiography data derived from large annotated ECG datasets (LUDB~\cite{LUDB} and ISP~\cite{ISP}) using the Kors transform~\cite{KORS1990}. 
Specifically, a U-Net with spatial and temporal attention was trained on ECG-derived vectorcardiography data to segment heartbeats into P-, QRS-, and T-waves. The transferability was evaluated in a zero-shot setting using averaged heartbeats derived from dipole domain reconstructions of the 31 fMCG recordings (cf. Section~\ref{ssec:cardiac_intervals}). An example of an automatically segmented fMCG vector loop and its individual channels is shown in Fig.~\ref{fig:delineation}. Performance metrics are summarized in Table~\ref{tab:delineation_fmcg}, while further implementation details are provided in \hyperref[a:delineation]{Supplement~7}. 

QRS detection achieved a near-perfect F1 score, demonstrating that the dominant cardiac feature transfers well. P-wave detection also yields high detection rates despite larger timing errors, likely due to annotation imprecision in the fMCG data and morphological differences between the ECG and MCG modalities. The T-wave detection proved more challenging, reflecting the larger domain shift for T-waves, which are typically very prominent in ECG but usually faint in fMCG \cite{strand2019WaveformCharacteristics}. T-wave-based analysis, therefore requires improved domain transfer. While the objective was to demonstrate transferability of the dipole representation rather than to build the best delineator, more advanced augmentation or fine-tuning on fMCG data would likely improve performance.

These results show that the dipole representation recovered by our forward model supports automated cardiac delineation on fMCG data. The successful transfer of ECG-trained models to fMCG demonstrates that this unified vectorcardiographic representation captures generalizable cardiac features, addressing the scarcity of annotated fMCG data.

\begin{table}[!t]
    \centering
    \caption{Automated Cardiac Delineation via Learned Vectorcardiographic Feature Transfer from ECG}
    \label{tab:delineation_fmcg}
    \setlength{\tabcolsep}{5pt}
    \begin{tabular}{ll|rrrrr}
        \toprule
        Wave & Point & SE (\%) & PPV (\%) & F1 (\%) & ME (ms) & Std (ms) \\
        \midrule
        P-wave & Onset  & 90.54 & 100.00 & 95.04 & $-21.28$ & 25.12 \\
        P-wave & Offset & 91.67 & 100.00 & 95.65 & $+4.08$  & 14.04 \\
        QRS    & Onset  & 98.81 & 100.00 & 99.40 & $-6.84$  & 8.20  \\
        QRS    & Offset & 98.81 & 100.00 & 99.40 & $-2.77$  & 9.56  \\
        T-wave & Onset  & 79.31 & 77.97  & 78.63 & $+69.35$ & 27.40 \\
        T-wave & Offset & 80.36 & 84.91  & 82.57 & $+0.89$  & 31.88 \\
        \bottomrule
    \end{tabular}
    \\[2pt]
    {\footnotesize SE: sensitivity, PPV: positive predictive value, ME: mean error, Std: standard deviation ($N=84$ beats). Detections within \SI{150}{ms} of the reference annotation are counted as true positives (ANSI/AAMI EC57:2012~\cite{ANSI_AAMI_EC57_2012}).}
\end{table}

\subsection{Reconstructed dipole positions}
\begin{figure}[!t]
    \centering
    \includegraphics[width=0.6\linewidth]{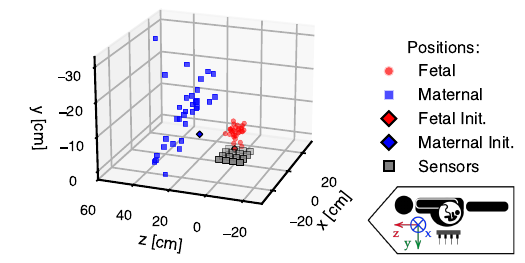}
    \caption{\textbf{Reconstructed fetal and maternal dipole positions across recordings.} Estimated dipole positions show clear separation between fetal and maternal cardiac sources.}
    \label{fig:3d_dipole_positions}
\end{figure}

In this section, we examine the reconstructed dipole positions for the $N=31$ measurements used in the cardiac interval evaluation (see Section~\ref{ssec:cardiac_intervals}). Fig.~\ref{fig:3d_dipole_positions} shows fetal and maternal dipole positions, their optimization initializations, and the sensor array in three-dimensional space. 

The estimated maternal dipole positions show greater spatial dispersion across recordings. This may be due to greater variation in maternal cardiac alignment, as sensor placement is optimized for fetal rather than maternal heart position, combined with a slightly higher uncertainty in maternal position estimation (see Fig.~\ref{fig:synthetic_fit}).
Nonetheless, maternal dipole positions remain separated from the fetal cluster, aligning with anatomical expectations. Fetal dipoles are tightly clustered near the sensor array, consistent with the abdominal placement over the array. Overall, the spatial distribution of reconstructed dipoles supports the physiological plausibility of the reconstructed sources.

\begin{figure}[!t]
    \centering
    \includegraphics[width=0.6\linewidth]{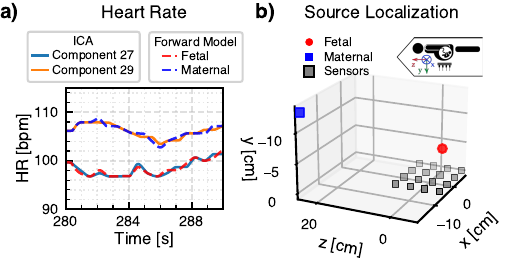}
    \caption{\textbf{Spatial localization enables fetal-maternal signal identification.}  \textbf{(a)} Fetal and maternal ICA components cannot always be distinguished by heart rate, since the fetal rate is not always higher. The forward model approach recovers the correct heart rate \textbf{(a)} and provides additional spatial information that supports source identification \textbf{(b)}.}
    \label{fig:ica_identification}
\end{figure}

In fMCG signal analysis, fetal and maternal signals cannot always be reliably distinguished by heart rate or signal amplitude. The maternal heart rate is not necessarily lower than the fetal heart rate, and signal strength varies with gestational age and the distance between sensors and sources.
Fig.~\ref{fig:ica_identification} shows an example where selecting ICA components based on heart rate is ambiguous and would lead to an incorrect assignment. 
In comparison, the forward model-based approach reconstructs sources in the dipole domain and estimates their source positions in three-dimensional space. This spatial information provides explicit details about the origin of cardiac activity and enables reliable distinction between fetal and maternal heart signals (see Fig.~\ref{fig:ica_identification} and Fig.~\ref{fig:3d_dipole_positions}). This represents a significant advantage over previous blind separation approaches that relied solely on field-domain analysis.

Although trained cardiologists may use additional physiological features, such as P- or T-wave prominence, to distinguish fetal from maternal signals, these criteria are complex heuristics and underscore that reliable ICA component selection is generally difficult to automate (see \hyperref[a:ica_limits]{Supplement~2} for additional discussion). In contrast, the forward model-based method facilitates fully automated, physiologically interpretable source reconstruction and identification.

\section{Discussion}

Reconstructing fMCG using a forward model shifts processing from statistical separation to physically constrained estimation.
The reconstructed cardiac vectors, analogous to vectorcardiography, provide a unified representation that incorporates all spatial directions and may be useful for standardization across subjects and gestational age. The estimated source positions separate between fetal and maternal activity, while position and orientation together may serve as a marker of fetal presentation or provide further anatomical insights. Validating this would require recordings with simultaneous reference measurements, which are currently not available.

As no ground truth for real fMCG exists, evaluation on real data is limited to agreement with reference methods rather than absolute accuracy. The used cohort consists only of healthy fetuses, so performance on pathological morphologies remains to be established.

The method uses a single equivalent dipole. Multipole or distributed-source models could, in principle, capture additional source structure, but they introduce challenges, including severe ill-posedness and computational complexity. Both multi-dipole and time-varying extensions would require strong fetal cardiac and movement priors, which currently do not exist.

Compact and cost-efficient systems could make fMCG more widely available, but it is still not a routine clinical practice. Automating signal reconstruction is a step towards automated fMCG workflows and supports potential clinical adoption, although diagnosis still requires experts. The forward model-based reconstruction also enables further technical developments in solving the inverse problem and integrating physiological priors.

\section{Conclusion}

In this paper, we present the first forward model-based reconstruction approach for fMCG, estimating dipole source signals in a physically motivated representation. 

Synthetic experiments with known ground truth confirm that the forward model recovers the underlying source activity. Validation on real fMCG data further shows strong agreement with manual spatial filtering-based processing ($\text{ICC} > 0.9$), and the estimated dipole positions are consistent with anatomical expectations.

The proposed framework enables automated fMCG reconstruction and returns source signals in a physiologically interpretable cardiac vector domain. This addresses key technical challenges in fMCG and supports its potential translation into prenatal care.

\printbibliography

% ===================================================================
% Supplementary Information
% ===================================================================
\clearpage
\setcounter{section}{0}
\setcounter{figure}{0}
\setcounter{table}{0}
\setcounter{equation}{0}
\renewcommand{\thesection}{S\arabic{section}}
\renewcommand{\thefigure}{S\arabic{figure}}
\renewcommand{\thetable}{S\arabic{table}}
\renewcommand{\theequation}{S\arabic{equation}}
\renewcommand{\theHsection}{S\arabic{section}}
\renewcommand{\theHfigure}{S\arabic{figure}}
\renewcommand{\theHtable}{S\arabic{table}}
\renewcommand{\theHequation}{S\arabic{equation}}

\phantomsection\pdfbookmark[0]{Supplementary Information}{supplement}
\begin{center}
{\Large\bfseries Supplementary Information for:\\
Automated Fetal Magnetocardiography Reconstruction Using a Biophysical Forward Model\par}
\end{center}
\vspace{1em}

\section{Preprocessing: Removing spatial and temporal outliers}\label{supp:preprocessing}

\subsection{Removing faulty sensor channels}
Faulty sensor channels exhibiting unphysical or outlying behavior are excluded prior to processing.
For each channel $c$ of sensor $s$, the 95-5 percentile interval is computed as 
\begin{equation}
    \mathcal{I}_{s,c} = \mathrm{P}_{95}(y_{s,c}) - \mathrm{P}_{5}(y_{s,c}),
\end{equation}
where $\mathrm{P}_{q}(\cdot)$ denotes the $q$-th percentile over time. Channels exhibiting signal ranges below the physical threshold $\tau_{\mathrm{phys}}$ (i.e., $\mathcal{I}_{s,c} < \tau_{\mathrm{phys}} = \SI{1}{pT}$) are excluded from subsequent analysis due to insufficient variability, indicating sensor failure and unphysical behavior.
For each valid channel, the temporal variability is measured by
\begin{equation}
    d_{s,c}(t) = \left\vert y_{s,c}(t+1) - y_{s,c}(t) \right\vert.
\end{equation}
Channels are flagged as outlying and removed if more than half of their samples fall outside the global variability range across all valid channels:
\begin{equation}
d_{s,c}(t) \notin \left[\mathrm{median}_s \mathrm{P}_{5}(d_{s,c}), \mathrm{median}_s \mathrm{P}_{95}(d_{s,c})\right].
\end{equation}
This robustly identifies both failed channels and sensors with abnormal temporal variability relative to the ensemble, addressing potential sensor malfunctions.

\subsection{Removing temporal artifacts}
To improve the robustness of the subsequent processing, temporal outliers and artifacts are removed using multi-resolution thresholding in the wavelet domain~\cite{strasserMotionArtifactRemoval2012}. A $7$-level stationary wavelet transform (SWT) with Haar wavelets is computed for each sensor channel $y_{s,c}$. Each wavelet coefficient sequence $c_j[n]$ at level $j$ is split into 10-second segments, for which the minimum and maximum values are determined. For each coefficient sequence, robust location $\hat\mu$ and scale $\hat\sigma$ estimates of the maximum and minimum are obtained using the median and normalized median absolute deviation (MADN). Thresholds are defined as $c_\mathrm{max} = \hat\mu_{\mathrm{max}} + 3\hat\sigma_{\mathrm{max}}$ and $c_\mathrm{min} = \hat\mu_{\mathrm{min}} - 3\hat\sigma_{\mathrm{min}}$ for each coefficient series. Samples lying outside these bounds are labelled as temporal artifacts. Finally, the fMCG recordings are divided into artifact-free segments of at least 10 seconds, excluding periods with substantial motion artifacts or magnetic interference.

\section{ICA annotation limitations}\label{a:ica_limits}

ICA-based reconstruction using Independent Component Subtraction requires manual inspection and annotation of all components into groups that are either retained or suppressed~\cite{yuMaternalMCGInterference2011}. This task is inherently challenging because ICA often fragments cardiac sources across multiple components. While fetal and maternal R-waves are usually distinct and prominent, identifying more subtle features, such as P- and T-waves, becomes increasingly difficult when they are distributed across components.

As a result, comprehensive annotation is highly laborious and often impractical, since subtle features may remain indistinguishable within noisy and residual components. In addition, ICA may fail to fully separate sources when their subspaces overlap and is sensitive to nonstationary signals~\cite{murtaSegmentedIndependentComponent2015}. 

\begin{figure}[!htb]
    \centering
    \includegraphics[width=\linewidth]{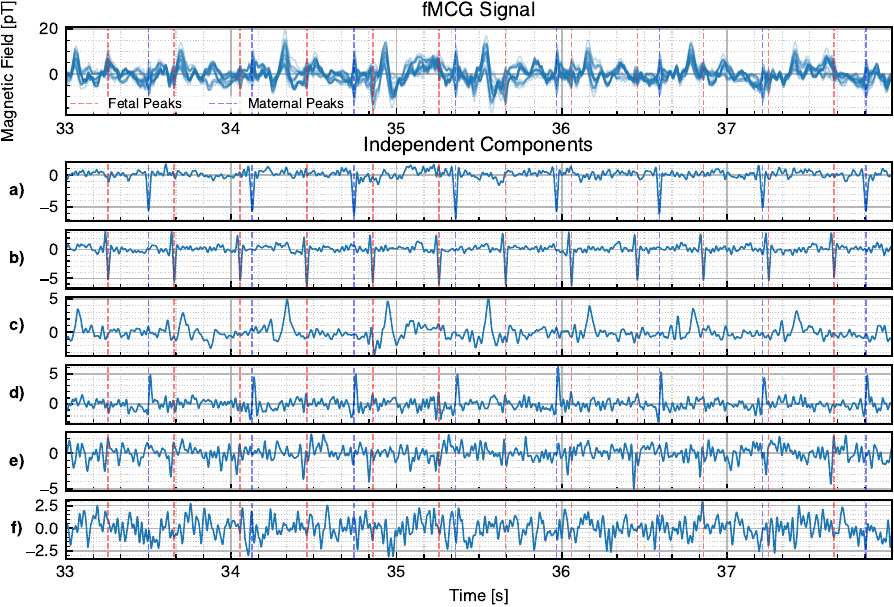}
    \caption{\textbf{fMCG recording and representative ICA components.} The magnetic field measurement is shown with all channels overlaid in the first plot. A selection of relevant ICA components is presented in \textbf{a)–f)}. The maternal \textbf{a)} and fetal \textbf{b)} R-peak series are clearly visible. Other cardiac features are often fragmented across components, such as the maternal T-wave in \textbf{c)}. Mixed components also occur, as shown in \textbf{d)}, where a predominantly maternal structure is present, but also minor fetal features at fetal peak positions are visible. Components \textbf{e)} and \textbf{f)} represent noisy mixtures with low signal-to-noise ratios, which complicate reliable assignment.}
    \label{fig:ica_fragmentation}
\end{figure}

Figure~\ref{fig:ica_fragmentation} illustrates this by showing the observed fMCG measurement after frequency filtering alongside relevant ICA components. While selecting the R-peak series is relatively straightforward, other cardiac components are fragmented across multiple components, making precise annotation nearly infeasible.

\section{Regularization}\label{a:regularization}

\begin{figure}[!htb]
    \centering
    \includegraphics[width=\linewidth]{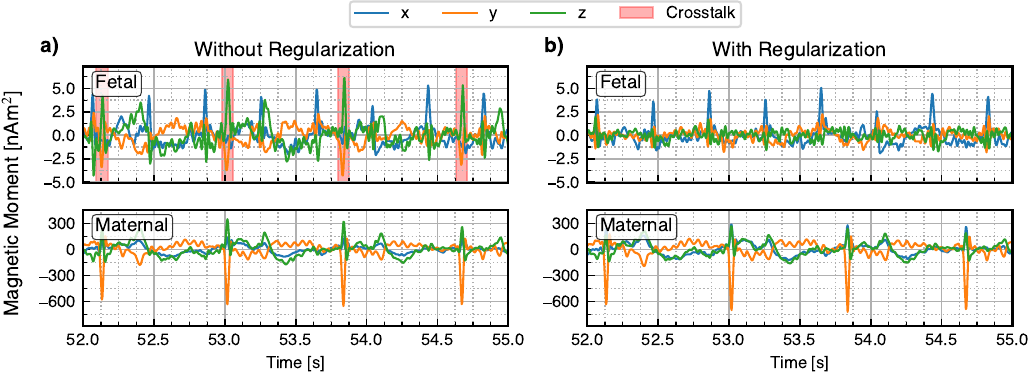}
    \caption{\textbf{Reconstructed magnetic dipole moments.} \textbf{a)} Without \textbf{b)} with cross-correlation regularization during optimization. Without regularization, notable overlap between fetal and maternal reconstructions is observed, while including the cross-correlation penalty improves the source separation.}
    \label{fig:reg_comparison}
\end{figure}

The effect of the cross-correlation regularizer during forward model reconstruction is shown in Figure~\ref{fig:reg_comparison} via the reconstructed magnetic dipole moments. Including the regularization term reduces overlap between both source components, thereby supporting source separation. 

    \subsection{Regularization strength}\label{ssec:regularization_strength}

    We investigate the sensitivity of the forward model reconstruction approach to the weight $\lambda$ of the cross-correlation regularization. We use the $N=31$ real fMCG recordings of the main text (gestational age: $21-40$ weeks, mean $\pm$ std: $31.1 \pm 4.98$) to analyze reconstructions under different regularization weights~$\lambda$. 
    A set of reference R-peak positions is obtained from manual ICA annotation by curation of independent components, yielding high-quality peak estimates using the visibility graph detector~\cite{emrich2023vg}.

    Let $a_{d_1 \mid d_2}$ denote the magnitude of the averaged field of source $d_1$ at R-peak locations of $d_2$, 
    \begin{equation}
        a^{\lambda}_{d_1 \mid d_2} = \left\lVert \frac{1}{|T_{d_2}|} \sum_{\tau \in T_{d_2}} \mathbf{y}^{\lambda}_{d_1}(\tau) \right\rVert,
    \end{equation}
    where $d_1,d_2 \in \{\text{fetal}, \text{maternal}\}$ and $\mathbf{y}^{\lambda}_{d_1}(t)$ is the field of source $d_1$, reconstructed using the regularization weight $\lambda$.

    We measure crosstalk as the ratio of one source's R-peak amplitude in another's reconstruction to its own R-peak amplitude. The maternal crosstalk in the fetal reconstruction is calculated as
    \begin{equation}
        X_{m \to f}(\lambda) = 20\log_{10}\frac{a^{\lambda}_{f \mid m}}{a^{\lambda}_{f \mid f}} \quad \text{dB}.
    \end{equation}
    With an analogous definition for fetal crosstalk in maternal reconstruction $X_{f \to m}$. Zero dB means the other source's R-peak amplitude equals its own source's R-peak amplitude, and lower values correspond to less crosstalk.

    In addition, two metrics track changes in the reconstructed waveforms. Amplitude change relative to no regularization ($\lambda = 0$) is computed as
    \begin{equation}
        \delta_f(\lambda) = 20\log_{10}\frac{a^{\lambda}_{f \mid f}}{a^{\lambda=0}_{f \mid f}} \quad \text{dB}
    \end{equation}
    and the R-peak detection performance, measured using the F1-score (defined in Section~\ref{ssec:eval}), serves as a proxy for signal quality and downstream performance.
    R-peak detection is performed on the reconstructed magnitude using the visibility-graph detector and compared against the manually curated R-peaks used for crosstalk analysis. 

\begin{figure}[!htb]
        \centering
        \includegraphics[width=\linewidth]{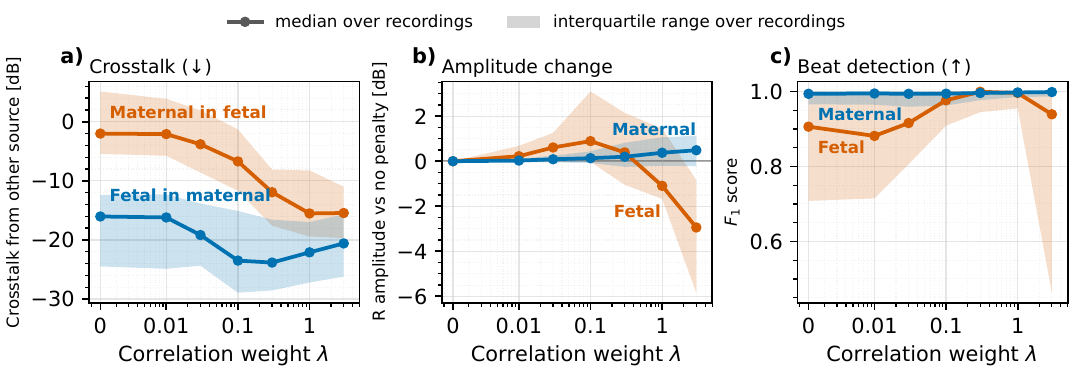}
        \caption{\textbf{Cross-correlation penalty weight ablation.} \textbf{a)} Crosstalk between the two sources, \textbf{b)} R-peak amplitude relative to no regularization, $\lambda=0$, \textbf{c)} R-peak detection performance against the curated annotation. Median and interquartile range over $N = 31$ recordings, fetal in orange and maternal in blue.}
        \label{fig:reg_ablation}
    \end{figure}

    Figure~\ref{fig:reg_ablation} shows crosstalk, amplitude change, and beat detection performance across regularization weights for both fetal and maternal reconstructions. Maternal crosstalk in fetal reconstruction is more prominent than the reverse. As the regularization weight increases, maternal leakage decreases monotonically. The fetal contribution in maternal components also decreases, but with a smaller effect, since the maternal signal is initially less contaminated by the much weaker fetal signal.
    The risk of overregularization is examined through amplitude change and beat detection performance. In the range $\lambda \in [0.01, 0.3]$, fetal amplitude increases slightly while maternal amplitude remains mostly unaffected. For larger values, fetal amplitude gradually decreases. A moderate amplitude reduction may be acceptable as long as fetal beats remain undistorted. This is measured by fetal beat detection performance, which improves from $F_1 = 0.906$ without regularization to $0.999$ around $\lambda = 0.3$ and drops again for $\lambda \gtrsim 1$. Maternal performance remains near perfect across all values.
    
    Cross-correlation regularization effectively reduces maternal leakage in the much weaker fetal reconstruction. For our cohort, we observe values within $[0.1, 0.5]$, providing a good balance, while smaller values tend to exhibit more crosstalk and larger values risk decreasing amplitude and detection performance. 

\section{Cardiac time intervals}\label{a:annotation_example}

In this work, cardiac time intervals were measured manually in the field domain from real recordings. 
Section~\ref{ssec:annotation} describes how the intervals were measured. Section~\ref{ssec:interval_ga} provides additional details to Section~\ref{ssec:cardiac_intervals} of the main text by stratifying the agreement between forward model based reconstruction and spatial filtering across gestational age.

\subsection{Annotation procedure}\label{ssec:annotation}
\begin{figure}[!htb]
    \centering
    \includegraphics[width=\linewidth]{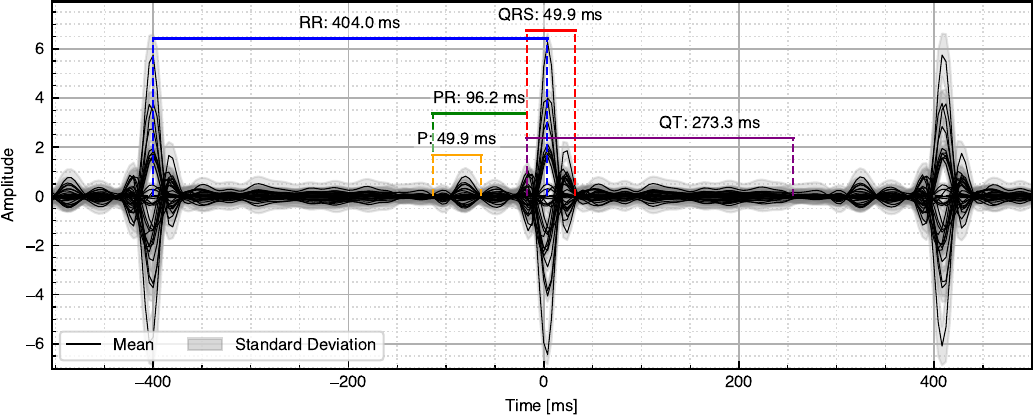}
    \caption{\textbf{Cardiac interval annotation example.} Averaged fetal heartbeat waveform reconstructed using the forward model and displayed as a butterfly plot in the field domain. Cardiac time intervals (P-wave, PR interval, QRS complex, and QT interval) are annotated at wave onsets and offsets.}
    \label{fig:annotation_example}
\end{figure}

Figure~\ref{fig:annotation_example} illustrates the manual annotation procedure for cardiac interval measurements used in Section~\ref{ssec:cardiac_intervals} of the main text. An averaged fetal heartbeat waveform is shown, depicting three consecutive beats with all sensor channels overlaid. 
Cardiac time intervals (P, PR, QRS, QT, and RR) are manually annotated in the field domain to enable direct comparison with spatial-filtering-based reconstructions.
The P-wave duration is measured from the onset to the end of the P~wave. The PR~interval extends from P-wave onset to QRS onset, reflecting the interval between atrial and ventricular depolarization. The QRS~interval measures ventricular depolarization, from the beginning to the end of the QRS complex. The QT~interval, from the onset of the QRS complex to the end of the T~wave, represents the full period of ventricular depolarization and repolarization. The RR~interval, measured between successive QRS peaks, indicates the time between ventricular beats. The corrected QT interval (QTc) was calculated using Bazett's formula $\text{QTc} = \text{QT}/\sqrt{\text{RR}}$.

    \subsection{Interval agreement across gestational age} \label{ssec:interval_ga}
    
\begin{figure}[!htb]
        \centering
        \includegraphics[width=\linewidth]{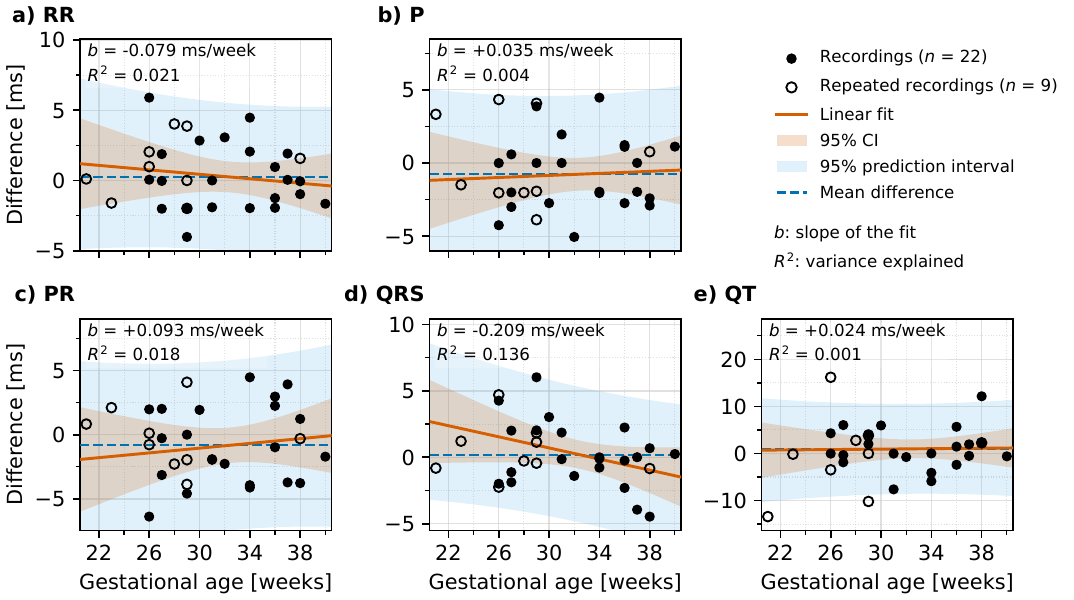}
        \caption{\textbf{Interval agreement across gestational age.} \textbf{a)} RR
        \textbf{b)} P \textbf{c)} PR \textbf{d)} QRS \textbf{e)} QT. Paired difference
        between the forward model and spatial filtering ($N = 22$, one recording per patient), with linear trend, its 95\% confidence interval, 95\% prediction interval, and the mean difference. $b$ is the fitted slope and $R^2$ the variance explained by gestational age. Additionally, open markers show the nine repeated recordings excluded from the paired analysis, giving the full cohort of $N = 31$.}
        \label{fig:ga_agreement}
    \end{figure}

    This section complements Section~\ref{ssec:cardiac_intervals} of the main text by investigating how the agreement between the forward model and spatial filtering varies across gestational age. Because this analysis requires independent observations, the fit uses one recording per patient $N=22$ (gestational age: $26-40$, mean $\pm$ std: $32.5 \pm 4.42$).    
    Figure~\ref{fig:ga_agreement} shows the paired difference in cardiac intervals between forward model reconstructions and spatial filtering reference, following Table~\ref{tab:reconstruction_diffs} in the main text. Open markers additionally show the nine repeated recordings excluded from the independent subset, giving the full cohort of $N=31$. For each cardiac interval, we regressed the differences between methods on gestational age using ordinary least squares. The slope $b$ gives the change in difference per week, and $R^2$ the ratio of variance explained by gestational age. QTc was excluded as its paired difference follows that of QT.
    
    For every interval, the mean difference lies within the 95\% confidence interval of the fitted linear trend. Gestational age explains less than 2\% of the variance for RR, P, PR, and QT, measured by $R^2$. QRS shows a slight trend ($R^2=0.14$), which amounts to \SI{-1.11}{ms} over the fitted gestational age, but remains below annotator variability.     
     The repeated recordings also lie within the 95\% prediction interval for all cardiac intervals except QT, where the exceptions are of opposite sign, indicating a larger spread rather than a trend.
     
    Spatial filtering is an established reference method for fMCG processing where no ground truth exists. Therefore, these results cannot establish the absolute accuracy of either method, but indicate that disagreement between the methods does not vary with gestational age.
    
\section{Relaxing the fixed dipole position constraint}\label{a:time_varying_positions}

Previously, cardiac dipole positions were approximated as fixed over time. While this simplification is reasonable for short time windows where the approximate stationarity holds, it may be relaxed for longer recordings or to capture fetal movement. In this section, we demonstrate that our forward model framework naturally extends to time-varying dipole positions $\mathbf{r}(t)$ and highlight the inherent limitation that additional regularization priors are required for stable reconstruction of movable dipoles.

\subsection{Piecewise constant parametrization and Total Variation regularization}
Allowing time-varying dipole positions introduces significantly more parameters to estimate, which makes the optimization problem harder and increasingly ill-posed without additional constraints.

To systematically explore this trade-off, we employ a piecewise constant parametrization where the recording is divided into temporal segments of fixed length, with dipole positions remaining constant within each segment. Varying the segment lengths allows us to control the degrees of freedom from fully constrained (fixed position) to increasingly unconstrained (many short segments). To further stabilize optimization, we apply a Total Variation (TV) regularizer that enforces smooth trajectories by penalizing rapid changes between consecutive segments:

\begin{equation}
    \mathcal{L}_{\text{TV}} = \frac{\lambda_{\text{TV}}}{6} \sum_{j=1}^{6} \sqrt{\frac{1}{N_{\text{diff}}}\sum_{i=1}^{N_{\text{diff}}} \left( [\mathbf{r}_{i+1}]_j - [\mathbf{r}_i]_j \right)^2}
\end{equation}
where $\lambda_{\text{TV}}$ controls trajectory smoothness, $\mathbf{r}_i \in \mathbb{R}^6$ is the stacked position vector of both dipoles at segment $i$, $[\mathbf{r}_i]_j$ denotes the $j$-component, and $N_{\text{diff}} = N_{\text{segments}} -1 $ is the number of consecutive differences. This normalization ensures that the term remains independent of recording length and the number of optimized parameters.

\subsection{Parameter sweep}
\begin{figure}[!htb]
    \centering
    \includegraphics[width=\linewidth]{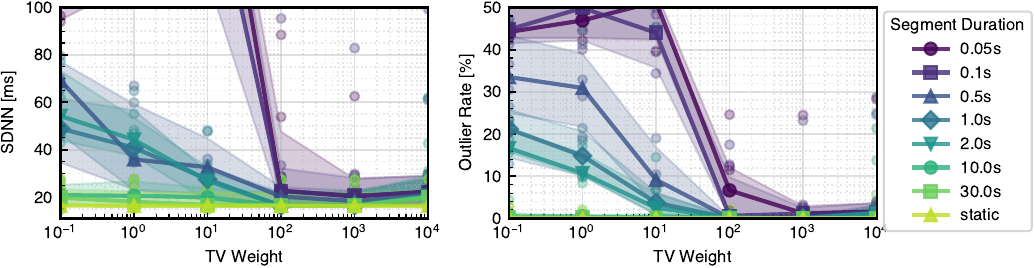}
    \caption{\textbf{Effect of TV regularization and constant segment duration on reconstruction quality.} With $\lambda_{\text{TV}} = 100$, short segment durations (\SIrange{0.5}{2}{\second}) achieve performance approaching the constant position baseline (static), while without regularization, increasing degrees of freedom lead to instability. Metrics shown are fetal outlier rate (lower is better) and SDNN (larger values indicate poorer heartbeat detection) computed from 12 patients.}
    \label{fig:tv_sweep_results}
\end{figure}

We evaluated a parameter grid on $N=12$ recordings, sweeping TV regularization weights and segment durations. Reconstruction quality was evaluated using automated heartbeat detection on the reconstructed signals, measuring fetal heart rate variability using SDNN and the outlier rate. Here, inflated SDNN serves as a proxy for misdetections in the automated heartbeat detection, and thus as an indicator of reconstruction quality. Figure~\ref{fig:tv_sweep_results} shows that with increasing degrees of freedom (decreasing segment length), performance degrades significantly without sufficient regularization. However, with reasonable TV regularization, e.g. $\lambda_{\text{TV}} = 100$, the moving-dipole reconstruction matches the static SDNN to within \SI{0.4}{\milli\second} per patient at {\SI{1}{\second}} segments, and performance remains comparable for segment durations as short as \SIrange{0.5}{2}{\second}.

This demonstrates that the framework allows reconstruction of fetal cardiac activity using movable dipoles in general, but dedicated movement priors would need to be developed and validated to achieve robust and comparable reconstruction quality.

\subsection{Reconstructed position trajectories}
\begin{figure}[!htb]
    \centering
    \includegraphics[width=\textwidth]{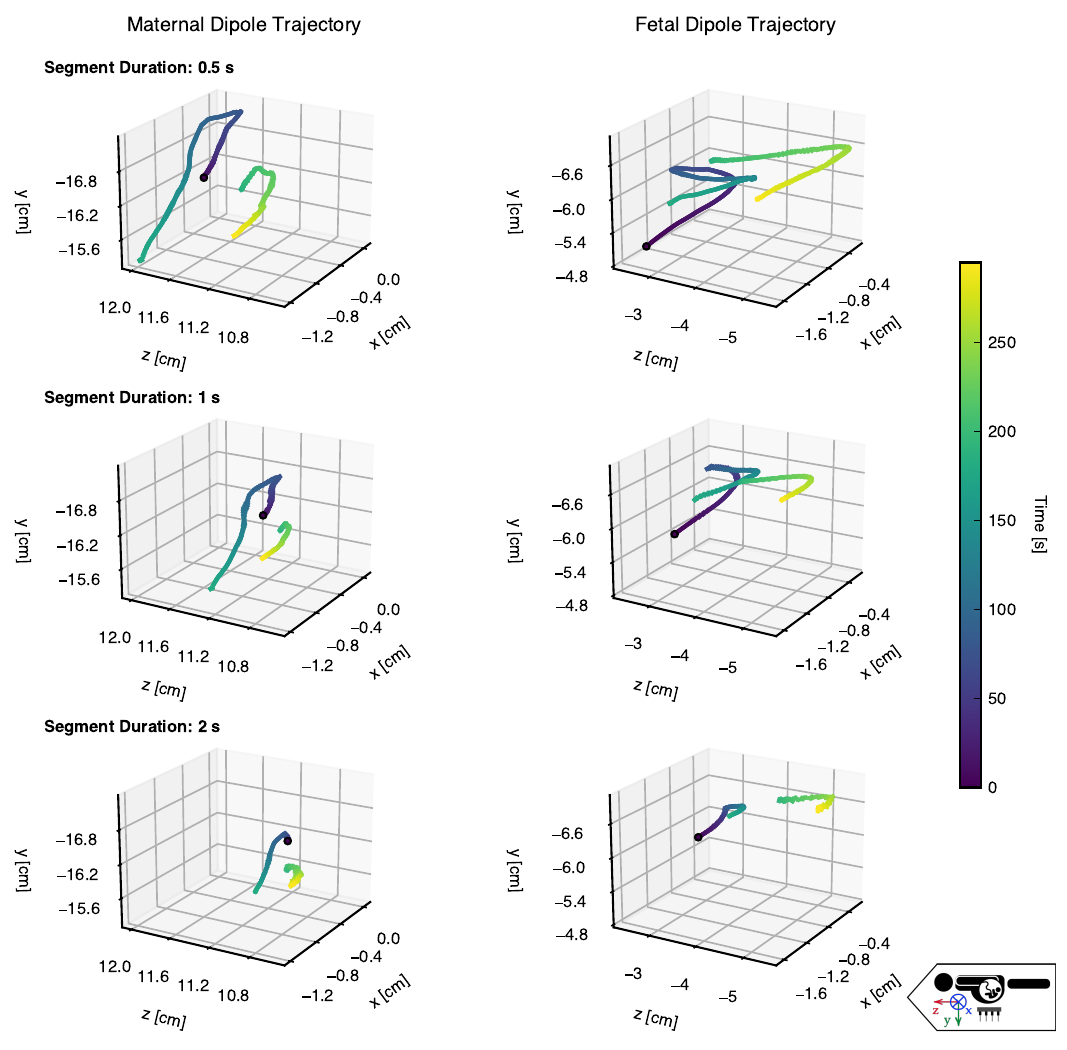}
    
    \caption{\textbf{Reconstructed 3D dipole position trajectories for varying segment lengths.} Fetal and maternal trajectories exhibit smooth and bounded motion within the sensor array geometry. Different regularization strengths, reflected in varying segment lengths, yield distinct levels of motion, while the overall patterns appear visually similar. Because temporal artifact segments were removed during preprocessing, the trajectory appears discontinuous.}
    \label{fig:3d_trajectories}
\end{figure}

Figure~\ref{fig:3d_trajectories} shows the reconstructed dipole position trajectories for $\lambda_{\text{TV}} = 100$ and $\SI{0.5}{s}, \SI{1}{s}$ and $\SI{2}{s}$ segment lengths for a representative measurement. 

The reconstructed trajectories exhibit smooth and spatially bounded motion within the sensor array geometry. 
However, we acknowledge that it remains unclear whether these trajectories represent true physical motion or rather optimization artifacts compensating for model mismatch. This ambiguity and dependence on the regularization constraint are emphasized, as different segment lengths, and thus regularization strengths, yield different trajectories. 

These observations motivate the development of more sophisticated movement priors tailored to fMCG. Validation would require comparison with independent fetal movement-tracking modalities. We defer such validation and the development of movement priors to future work, but highlight that our framework naturally enables this direction. For the cardiac interval analysis in the main text, the constant position model provides a sufficient and interpretable baseline. However, the variability of estimated trajectories across regularization settings raises the question of whether this positional ambiguity also affects the clinically relevant cardiac waveform outputs.

\subsection{Clinical output robustness under positional ambiguity}
To investigate whether positional ambiguity translates into instability of the cardiac outputs, synthetic fMCG recordings with known ground truth were simulated. Signals were generated using the dipole model with positions following slowly time-varying trajectories modeled as Ornstein-Uhlenbeck processes, fluctuating within {\SI{\pm5}{\milli\meter}} per axis at a mean speed of {\SI{2}{\milli\meter/\second}}. Cardiac moments were derived from real vectorcardiographic recordings. Real fMCG sensor noise was added at \SI{0}{dB} signal-to-noise ratio, combined with \SI{5}{\percent} Gaussian noise. Reconstruction was performed using 1-second piecewise constant position segments with varying TV regularization weights. Position, moment, and SDNN errors were evaluated against ground truth across $N=12$ synthetic realizations. 

\begin{figure}[!htb]
    \centering
    \includegraphics[width=\linewidth]{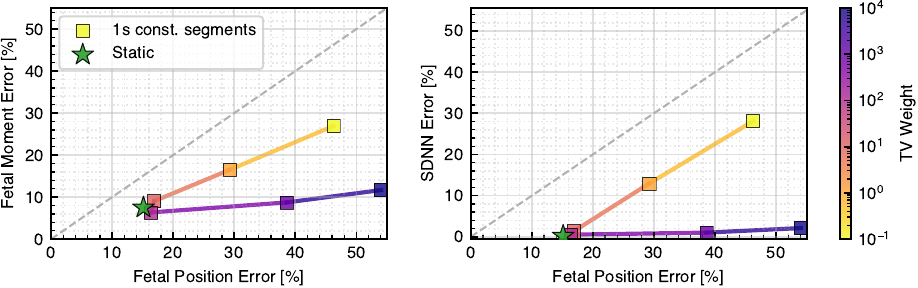}
    \caption{\textbf{Position--moment tradeoff on synthetic data.} \textbf{(a)} Position error versus moment error for varying TV regularization weights (color) using 1-second constant segments. \textbf{(b)} Position error versus SDNN error. Under moderate TV weights, moment and SDNN errors remain low and comparable to the static dipole baseline (star) despite position variability, supporting the static dipole assumption adopted in the main text.}
    \label{fig:trade_off}
\end{figure}

Figure~\ref{fig:trade_off} shows fetal magnetic moment and SDNN errors versus position error across regularization settings. 
Decreasing TV weight allows greater position freedom but increases moment error, whereas excessive regularization leads to overly smoothed trajectories. With moderate TV weights, both errors can be minimized simultaneously, achieving performance comparable to the static dipole baseline (star). At $\lambda_\text{TV}=100$, fetal moment error is \SI{6.4}{\percent} for TV regularized versus \SI{7.5}{\percent} for the static baseline, while SDNN error is \SI{0.5}{\percent} versus \SI{0.2}{\percent}, respectively.

This demonstrates that cardiac waveform reconstruction remains stable despite position variability across regularization settings. Moreover, the fixed dipole assumption represents a reasonable trade-off that achieves a comparable level of reconstruction accuracy without requiring complex motion priors. 
Extending to time-varying dipole positions would additionally require sophisticated movement priors and validation against independent fetal movement tracking modalities, which remain open questions for future work.

\section{Optimization details}\label{a:optimization}

The optimization parameters are partitioned into three groups, $\mathbf{\theta} = \{ \mathbf{\theta}_{\mathbf{m}_f}, \mathbf{\theta}_{\mathbf{m}_m}, \mathbf{\theta}_{\mathbf{r}} \}$ comprising the fetal dipole moment over time $\mathbf{\theta}_{\mathbf{m}_f} = {\mathbf{m}_{f}(t)}|_{t=0}^{T-1}$, maternal dipole moment over time $\mathbf{\theta}_{\mathbf{m}_m} = {\mathbf{m}_{m}(t)}|_{t=0}^{T-1}$ and dipole positions $\mathbf{\theta}_{\mathbf{r}} \in  \mathbb{R}^6$, enabling group-wise gradient normalization to compensate for differing magnitudes across temporal and spatial parameters.

\paragraph{Parameter normalization and scaling}
Prior to optimization, the observed field is normalized, and the optimization parameters $\mathbf{m}$ and $\mathbf{r}$ are rescaled based on initial estimates and prior knowledge of their expected magnitudes, mapping all optimization parameters to comparable scales.

\paragraph{Parameter initialization}
Initializations for the dipole positions $\mathbf{r}$ are found by selecting arbitrary values within physiological plausible regions, while the dipole moments are initialized using a robust pseudo-inverse estimate of the whitened forward model: $\mathbf{m}_\mathrm{init} = \mathbf{F}^\dagger_\varepsilon(\mathbf{r}_\mathrm{init}) \, \mathbf{y}_W$, where $\mathbf{F}^\dagger_\varepsilon$ is the truncated Moore-Penrose pseudo-inverse of the whitened forward operator, discarding singular values smaller than $\varepsilon \sigma_{\max}$. A large threshold $\varepsilon$ ensures a stable and robust initialization.

\paragraph{Bounded reparametrization}
To ensure dipole locations within physically and physiologically plausible regions, the dipole position vector $\mathbf{r}$ is reparametrized using a bounded, differentiable, element-wise mapping:
\begin{equation}
    \mathbf{r}_{\text{bound}} = \xi(\mathbf{\theta}_{\mathbf{r}}) = \mathbf{\gamma}_0 + (\mathbf{\gamma}_1 - \mathbf{\gamma}_0) \sigma(\mathbf{\theta}_{\mathbf{r}}),
\end{equation}
where $\mathbf{\theta}_{\mathbf{r}}$ are the unconstrained optimization parameters of the dipole location, $\mathbf{\gamma}_0,\mathbf{\gamma}_1 \in \mathbb{R}^6$ are the per-axis per-dipole lower and upper bounds, and $\sigma(\cdot)$ denotes the logistic sigmoid function applied element-wise, defined as $\sigma(\mathbf{\theta}) = \frac{1}{1+e^{-\mathbf{\theta}}}$.
This maps the optimization space of $\mathbf{r}$ effectively into a bounded and differentiable parameter space, with spatial boundaries for each parameter and restricts the solution to physically meaningful regions within the fMCG device, thereby eliminating ambiguities from non-physical symmetries in the solution space, e.g., dipoles located within the sensor array or beneath the patient bed.
The boundary values $\mathbf{\gamma}_0,\mathbf{\gamma}_1$ are determined empirically based on the fMCG device's dimensions and physiological constraints. 

\paragraph{Optimizer selection}
While other optimization methods, such as Adam~\cite{Kingma2015Adam}, can also be utilized to solve this problem, we found that Lamb's groupwise normalization can improve the reconstruction performance in certain scenarios, particularly in compensating for the differing orders of magnitude and loss landscapes across fetal and maternal dipole moments and spatial coordinates.

\paragraph{Hyperparameters}
All experiments used $500$ optimization iterations, with initial learning rates of $10^{-1}$ for dipole moments and $10^{-2}$ for positions. Learning rates were reduced when the loss began to plateau.
A warmup phase of $50$ steps was applied prior to optimization, which stabilizes training and helps initialize the optimizer's state. The temporal correlation penalty was computed using 10-second windows, a strength of $\lambda = 5 \cdot 10^{-1}$ and time lags within $\pm \SI{100}{ms}$ (spanning typical atrio-ventricular conduction times).
The dipoles were initially positioned at $\mathbf{r}_{f, \text{init}} = [0, -2, 0]~\mathrm{cm}$ and $\mathbf{r}_{m, \text{init}} = [0, -5, 20]~\mathrm{cm}$ in the $(x, y, z)$ coordinate system.

Reasonable variations in these hyperparameters yield qualitatively similar results, while longer optimization primarily improves data fitting without substantially altering the reconstruction quality. Section~\ref{ssec:regularization_strength} provides an ablation of the cross-correlation weight $\lambda$.

\section{Automated cardiac delineation via domain transfer}\label{a:delineation}

To demonstrate the utility of our three-dimensional dipole representation, we trained a 1D U-Net to segment cardiac waves (P-wave, QRS complex, T-wave) on vectorcardiography derived from large annotated ECG datasets. This approach circumvents the scarcity of annotated fMCG data and validates transferability from the ECG domain to our dipole source representation.

\subsection{Data}
The training data comprised two large annotated ECG datasets: LUDB~\cite{LUDB} (200 recordings with fiducial point annotations) and ISP~\cite{ISP} (475 recordings with segmentation masks), both of which provided 10-second 12-lead ECG measurements. LUDB annotations were converted to dense segmentation masks in consistency with ISP.

Vectorcardiographic representations were derived using the Kors transformation~\cite{KORS1990}, converting 8 ECG leads (I, II, V1-V6) into three orthogonal components (x, y, z). This transformation preserves temporal annotations while providing a spatial representation analogous to our dipole source domain.

\subsection{Model architecture}
We implemented a 1D U-Net with spatial and temporal attention (38.9M parameters) for 3-channel VCG input, producing 4-class segmentation outputs (background, P-wave, QRS, T-wave).

The encoder consists of four downsampling blocks with channels $\{64, 128, 256, 512\}$ and Conv1D kernel sizes $\{9, 7, 5, 3\}$, each followed by MaxPool1D (stride 2).

The bottleneck applies axial attention, comprising 4-head self-attention for temporal (sinusoidal positional encoding) and spatial (learnable embeddings) components. Both modules include dropout ($p=0.3$) and $4\times$ feedforward expansion.

The decoder mirrors the encoder with transposed convolutions (kernel 4, stride 2) and skip connections. A final $1 \times 1$ convolution produces per-timestep class logits.

\subsection{Data preprocessing \& augmentation}
All ECG recordings were resampled to \SI{500}{\hertz} and preprocessed with \SI{50}{\hertz} powerline notch filtering, followed by FIR bandpass filtering (\SIrange{0.67}{45}{\hertz}). Signals were normalized to a range of $[-1, 1]$ to preserve spatial relationships between leads and segmented into \SI{2}{\second} windows.

During training, we applied the following data augmentation with \SI{80}{\percent} probability per technique to improve model robustness and support domain transfer to fMCG signals:
\begin{enumerate}
    \item Temporal scaling (\SIrange{40}{110}{\percent} speed): varies heart rate
    \item Temporal shifting ($\pm$\SI{1}{\percent}): shifts sample positions within window
    \item Gaussian noise ($\sigma = 0.2 \times \sigma_\text{signal}$): simulates measurement noise
    \item Amplitude scaling (\SIrange{90}{110}{\percent}): varies signal amplitude
    \item Baseline wander (\SIrange{0.1}{16}{\hertz} sinusoids): simulates drift artifacts
    \item 3D rotation ($\pm$\SI{180}{\degree} on all axes): varies spatial orientation
    \item Realistic ECG artifacts: simulates motion, muscle, and powerline interference
\end{enumerate}
Augmentations were applied before final normalization, with temporal augmentations modifying both signals and annotations.

\subsection{Training}
The model was trained for 100 epochs with a batch size of 64, using focal loss ($\gamma=2.0$) and class weights determined from the empirical class distributions in the training set to account for class imbalance.

The training and validation sets comprised 1,373 and 295 two-second windows, respectively (70\%/15\% split), maintaining balanced proportions from the LUDB and ISP datasets. We used the AdamW optimizer with a learning rate of $10^{-3}$ and weight decay of $10^{-4}$, both determined through hyperparameter tuning. The learning rate was reduced by half when the validation performance plateaued (after 5 epochs of patience), with gradient clipping set to 1.0 and early stopping (with a patience of 20 epochs).

\subsection{Evaluation}\label{ssec:eval}
While the model was trained on sample-wise annotation masks, evaluation was performed on fiducial points (wave onsets and offsets) to align with standard delineation benchmarks.
Predictions within a \SI{150}{ms} tolerance window (ANSI/AAMI EC57:2012 standard~\cite{ANSI_AAMI_EC57_2012}) of ground truth annotations were counted as true positives (TP), while unmatched predictions and unmatched annotations were counted as false positives (FP) and false negatives (FN), respectively. 

The following detection metrics were computed:
\begin{equation}
    \text{Sensitivity (SE)} = \frac{\text{TP}}{\text{TP} + \text{FN}} \times 100\%
\end{equation}
\begin{equation}
    \text{Positive predictive value (PPV)} = \frac{\text{TP}}{\text{TP} + \text{FP}} \times 100\%
\end{equation}
\begin{equation}
    \text{F1-score} = 2 \cdot \frac{\text{SE} \cdot \text{PPV}}{\text{SE} + \text{PPV}}
\end{equation}

Timing accuracy was quantified as:
\begin{equation}
    \text{Mean Error (ME)} = \frac{1}{N_{\text{matched}}} \sum_{i=1}^{N_{\text{matched}}} \left( t_i^{\text{pred}} - t_i^{\text{ref}} \right)
\end{equation}
\begin{equation}
    \text{Standard Deviation (SD)} = \sqrt{\frac{1}{N_{\text{matched}}} \sum_{i=1}^{N_{\text{matched}}} \Big( (t_i^{\text{pred}} - t_i^{\text{ref}}) - \text{ME} \Big)^2 }
\end{equation}
where $t_i^{\text{pred}}$ and $t_i^{\text{ref}}$ are the predicted and reference fiducial times for the $i$-th matched detection, and $N_{\text{matched}}$ is the total number of matched points. These metrics capture both the detection performance and the temporal precision of the fiducial points.

\subsection{Performance on ECG}
On the held-out ECG test set (295 two-second windows), the model achieved near-perfect performance across all cardiac waves (see Table~\ref{tab:ecg_fiducial_performance}), closely matching state-of-the-art ECG delineation approaches \cite{LUDB, emrich_prominence_24}. All wave boundaries except the T-wave offset were detected perfectly (F1-score = \SI{100}{\percent}, no false positives), while the T-wave offset achieved an F1-score of \SI{99.92}{\percent}. Timing accuracy was high, with low mean errors, and standard deviations well within the recommendations of the committee for Common Standards for Electrocardiography (CSE) \cite{cse1985recommendations}, i.e., the standard deviation of the mean error should be less than $2\sigma_{\text{CSE}}$ (see last column Table~\ref{tab:ecg_fiducial_performance}), a robust estimate of the error variability expected from expert cardiologists based on measurements made on different experts annotations.

Consequently, the trained delineator successfully learned the vectorcardiographic features from ECG data.

\begin{table}[hbt]
    \centering
    \caption{Fiducial point detection performance on the ECG test set with a \SI{150}{ms} tolerance.}
    \label{tab:ecg_fiducial_performance}
    \small\setlength{\tabcolsep}{4pt}
    \begin{tabular}{ll|rrrrrrrr|r}
        \toprule
        Wave & Point & TP & FP & FN & SE (\%) & PPV (\%) & F1 (\%) & ME (ms) & Std (ms) & $2\sigma_{\text{CSE}}$ \\
        \midrule
        P-wave & Onset & 585 & 0 & 0 & 100.00 & 100.00 & 100.00 & -2.07 & 2.68 & 10.2\\
        P-wave & Offset & 584 & 0 & 0 & 100.00 & 100.00 & 100.00 & +2.05 & 3.20 & 12.7\\
        QRS & Onset & 701 & 0 & 0 & 100.00 & 100.00 & 100.00 & -1.61 & 2.98 & 6.5\\
        QRS & Offset & 699 & 0 & 0 & 100.00 & 100.00 & 100.00 & +1.87 & 2.95 & 11.6\\
        T-wave & Onset & 654 & 0 & 0 & 100.00 & 100.00 & 100.00 & -2.37 & 7.54 & - \\
        T-wave & Offset & 662 & 0 & 1 & 99.85 & 100.00 & 99.92 & +1.50 & 5.42 & 30.6\\
        \bottomrule
    \end{tabular}
\end{table}

\paragraph{Transfer to fMCG}
For fMCG evaluation, the ECG-trained model was applied directly via zero-shot inference on a subset of $N=31$ recordings. Results are shown in Figure~\ref{fig:delineation} of the main text.
The successful transfer of learned cardiac features from vectorcardiography to our dipole domain highlights the inherently shared characteristics and demonstrates the feasibility of bridging ECG information to fMCG.

\end{document}